\documentclass[twocolumn]{autart}
\usepackage{amsmath}
\usepackage{amssymb} 
\usepackage{mathtools}
\usepackage{pstool}
\usepackage{subcaption}
\usepackage{subdepth}

\usepackage[dvipsnames]{xcolor}
\usepackage{soul} 
\renewcommand{\rm}[1]{{\color{black} #1}}
\newcommand{\bm}[1]{{\color{black} #1}}

\begin{document}

\begin{frontmatter}

\title{String stable integral control design for vehicle platoons with disturbances}
\author[QUT]{Guilherme Fr\'{o}es Silva}\ead{g.froessilva@qut.edu.au},
\author[UON]{Alejandro Donaire}\ead{alejandro.donaire@newcastle.edu.au},
\author[QUT]{Aaron McFadyen}\ead{aaron.mcfadyen@qut.edu.au},
\author[QUT]{Jason Ford}\ead{j2.ford@qut.edu.au}

\address[QUT]{School of Electrical Engineering and Robotics, Queensland University of Technology, 2 George St, 4000, QLD, Australia}
\address[UON]{School of Engineering, University of Newcastle, University Drive, 2308, NSW, Australia}


\begin{abstract}
 \bm{This paper presents a control design with integral action for vehicle platoons with disturbance that ensures string stability of the closed loop and disturbance rejection.}
 The addition of integral action and a coordinate change allows to develop sufficient smoothness conditions on the closed loop system to ensure that the \bm{closed loop system using the proposed controller is string stable} in the presence of time-varying disturbances and able able to reject constant disturbances. In addition, bounds for the tracking error of the platoon configuration are also given.  Further, a case study is considered together with a suitable controller structure, which satisfies the required smoothness conditions. Simulation results illustrate the performance of the closed loop.
\end{abstract}

\end{frontmatter}

\section{Introduction}
Multi-agent systems are common in natural, social and engineering systems, including cooperative systems \cite{Arcak2007,Knorn2016a,YangQuanChen2005}, intelligent transportation system \cite{Wit1999,Oncu2011}, and aerial vehicles coordination \cite{Cai2011,Pant2002}. 
Due to advantages of increasing traffic throughput and reducing fuel consumption \cite{Naus2010}, vehicle platooning has received recent attention, see \cite{Bergenhem2012} and references within. Generally, each platoon agent is coordinated to maintain a desired distance, or time headway, to its neighbours \cite{Seiler2004}. 

Research on vehicle platoons and intelligent transportation systems dates back to 1960s \cite{Levine1966}, which introduced the idea of controlling vehicle positions based on sensor and communication data. It was then realised that additional stability properties are desirable for vehicle platoons \cite{Peppard1974}, in particular string stability. A platoon system is said to be string stable when disturbances and initial conditions are attenuated along the string, regardless of the string length \cite{Swaroop1996,Besselink2017a}.

The communication structure, spacing policy, and vehicle dynamics of platoons impact their behaviour and stability properties, see \cite{Feng2019,Studli2017} and references within. The communication structure can be, among other topologies, unidirectional, where information travels only one way in the string \cite{Pant2002,Rogge2008}, and bidirectional, in which information travels both ways along the string \cite{Seiler2004,Arcak2007,Knorn2014,Herman2017}. Furthermore, bidirectional strings can be divided into symmetric or asymmetric strings according to the symmetric or asymmetric coupling between preceding and following vehicles \cite{Herman2017,Monteil2019}. Spacing policies, in turn, determine the desired distance between vehicles, which can be a function of the vehicle speed \cite{Darbha1999}. The spacing policy affects the system stability and agents behaviour. It is shown in \cite{Seiler2004,Barooah1997} that strings using only relative spacing information with constant spacing policy, and a double integrator model for the vehicle dynamics, are always $L_2$ string unstable for any linear controller.  In general, vehicle dynamics are considered linear \cite{Cook2005,Eyre1998,Pant2002,Rogge2008}, which allows the use of transfer functions and $H_\infty$ system norm string stability to study $L_2$ string stability. Nonlinear methods have been used in \cite{Knorn2014,Ferguson2017,Monteil2019}, which are results in control system with better performance. 

In the weaker $L_2$ string stability setting, an alternative to $L_\infty$ string stability of \cite{Swaroop1996}, it has been shown that symmetry in position coupling combined with asymmetry in velocity coupling improves the performance of heterogeneous, asymmetric, bidirectional platoons \cite{Herman2017}. An extension of string stability that considered disturbances has been proposed in \cite{Besselink2017a}. Recently, sufficient conditions that ensures disturbance string stability (DSS) were proposed in \cite{Monteil2019}. These conditions were used for control design, however, the presence of nonzero mean disturbances produces a nonzero steady-state error.

\bm{The key contribution of this paper is to propose a control design with integral action that provides disturbance string stability properties for a bidirectional platoon of vehicles with constant spacing policy subject to time varying disturbances. In addition, the proposed controller rejects constant disturbances. We use DSS sufficient conditions to obtain the controller gains, and then evaluate the performance of the proposed controller in simulation. Thus, we extend the results in \cite{Monteil2019} by considering a dynamic compensation of constant disturbances while ensuring DSS of the interconnected system. }

The remaining of the paper is organised as follows. We introduce the notation and formulate the problem in Section \ref{sec:Problem}. The control design and sufficient conditions for string stability are discussed in Section \ref{sec:Controller}. We present a numerical example to illustrate the control system performance in Section \ref{sec:Results}. We wrap up the paper with the conclusions in Section \ref{sec:Conclusion}.

\bm{\section{Notation and Problem Formulation} \label{sec:Problem}
\subsection{Notation}
Let $x$ be a vector in $\mathbb{R}^n$ and consider the signal $s(\cdot)\colon\mathbb{R}_+\to\mathbb{R}^n$. We define $\left\lvert x \right\rvert_2$ as the $L_2$ vector norm of $x$, and the $L_\infty$ function norm $\left\lVert s(\cdot) \right\rVert_{\infty} = \sup_{t}\left\lvert s(t) \right\rvert_{2}$. We let $A$ be a matrix in $\mathbb{R}^{n\times n}$ and define the $L_2$ matrix norm $\left\lVert A \right\rVert_2$ and the matrix measure $\mu_2(A)$, induced by the $L_2$ norm, as $\mu_2(A) = \max_i\left(\lambda_i[A]_s\right)$ and $\left[A\right]_s$ as the symmetric part of $A$, where $\lambda(A)$ are the eigenvalues of $A$. Additionally, the minimum and maximum singular value of the matrix $A$ are $\sigma_{\min}(A)$ and $\sigma_{\max}(A)$. A function $\Gamma(\cdot)\colon R_+\to R_+$ is of class-$\mathcal{K}$ if it is strictly increasing and $\Gamma(0)=0$, and of class-$\mathcal{L}$ if it monotonically decreases to 0 as its parameter goes to infinity. Finally, a function $\gamma(\cdot,\cdot)$ is of class-$\mathcal{KL}$ if $\gamma(\cdot,t)$ is a class-$\mathcal{K}$ function $\forall t\geq0$ and $\gamma(a,\cdot)$ is a class-$\mathcal{L}$ function $\forall a\geq 0$.

\subsection{Problem Formulation}}
We consider an interconnected system composed of $N\geq1$ agents whose dynamics can be described as follows
\begin{equation}
	\begin{split}
	    \dot{x}_{1i} &= f_{1i}(x_{1i},x_{2i}) \\
		\dot{x}_{2i} &= f_{2i}(x_{1i},x_{2i}) + u_i + d_i,
	\end{split}
	\label{eq:sys}
\end{equation}
for all $i=\{1,\dots, N\}$, with $x_{1i},x_{2i} \in \mathbb{R}^n$, $u_i \in \mathbb{R}^n$ and $d_i \in \mathbb{R}^n$, where $n\geq1$. The state vector, control input and the disturbance of the $i$th vehicle are $x_i = [x_{1i}^T \; x_{2i}^T]^T$, $u_i$ and $d_i$ respectively. The disturbances has two components $d_i = w_i(t)+\bar{w}_i$, where $w_i(t)$ and $\bar{w}_i$ are the time varying and constant components of the disturbance. The smooth functions $f_{1,2i} \colon \mathbb{R}^n \times \mathbb{R}^n\to \mathbb{R}^n$ describe the system dynamics $f_i(x_i) = [f_{1i}^T \; f_{2i}^T]^T$. We define a virtual vehicle as the reference state $x_0$.

A desired property for interconnected vehicle systems is the string stability property, which will ensure for example that disturbances are not amplified when they propagate along the string while maintaining a desired configuration. We express the control objective requiring that the state $x_i$ should converge to the desired configuration $x_i^\star$, where $x_i^\star$ verifies $\dot{x}^\star_i=f_i(x_i^\star)$, as well as it is a solution of the system in the absence of disturbances.

\bm{A definition of string stability expressed in $\varepsilon$--$\delta$ form was proposed by Swaroop and Hedrick \cite{Swaroop1996} and implies that the state deviations from the origin are not amplified along the platoon. In \cite{Besselink2017a}, Besselink and Johansson proposed the concept of disturbance string stability to capture the effects of external disturbances. This definition is an extension of classical string stability but expressed in term of class-$\mathcal{K}$ and class-$\mathcal{KL}$ functions. In this paper, we use the disturbance string stability definition in \cite{Besselink2017a}.}

\begin{defn}[Disturbance String Stability]
	\bm{Consider the system \eqref{eq:sys}	and assume that $x_i^\star$ is a solution to its unperturbed dynamics. Then, the equilibrium $x_i^\star$ is said to be \emph{disturbance string stable} if there exists a $\mathcal{KL}$ function $\gamma$ and a $\mathcal{K}$ function $\Gamma$ such that, for any disturbance $d_i$ and initial conditions the estimate}
	\begin{align}\label{eq:estimatedss}
	\begin{split}
	\sup_i\left\lvert x_i(t)-x_i^\star(t) \right\rvert_2&\leq\gamma\left(\sup_i\left\lvert x_i(0)-x_i^\star(0) \right\rvert_2,t\right)\\&+\Gamma\left(\sup_i\left\lVert d_i(t) \right\rVert_{\infty}\right)
	\end{split}
	\end{align}
is verified for all $t>0$ \bm{and $N\in \mathbb{N}$. The functions $\gamma(\cdot,\cdot)$ and $\Gamma(\cdot)$ are the same for any platoon length $N$, and thus the estimate of the state error norm is independent of the number of agents.} 
	\label{def:ss_baselink}
\end{defn}

\bm{Notice that \eqref{eq:estimatedss} ensures  asymptotic stability for the undisturbed case (see \cite[Section 2.5]{Sontag2008}), which implies $x_i\to x_i^\star$ as $ t \to \infty$.}

We propose the following form for the controller $u_i$
\begin{align}
\begin{split}
u_{i} &= h_{i,i-1}(t,x_{i},x_{i-1}) + \varepsilon_i h_{i,i+1}(t,x_{i},x_{i+1}) \\ & +
h^{0}_{i}(t,x_{i},x_0) + k\zeta_i 
\label{eq:controller_ac}
\end{split}\\
\begin{split}
\dot{\zeta_i} &= g_{i,i-1}(t,x_{i},x_{i-1}) + \varepsilon_i g_{i,i+1}(t,x_{i},x_{i+1}) \\ &+ g^{0}_{i}(t,x_{i},x_0)	
\end{split}
\label{eq:integraldynamics}
\end{align}
where the smooth functions  $h_{i,j}\colon\mathbb{R}_+\times \mathbb{R}^{2n} \times \mathbb{R}^{2n} \to \mathbb{R}^n$ represent the coupling functions between neighbour vehicles $i$ and $j$, while $h^0_{i}\colon\mathbb{R}_+\times \mathbb{R}^{2n} \times \mathbb{R}^{2n} \to \mathbb{R}^n$ is the coupling of vehicle $i$ with the reference state $x_0$, and $\varepsilon_i \in [0,1]$ is the symmetry constant for vehicle $i$, which weights its coupling with the following vehicle. The controller state $\zeta \in \mathbb{R}^n$ allows for integral action to compensate disturbances. The constant $k$ is the integral action gain and the smooth functions $g_{i,j}\colon\mathbb{R}_+\times \mathbb{R}^{2n} \times \mathbb{R}^{2n} \to \mathbb{R}^n$ shape the integral action dynamics \eqref{eq:integraldynamics}.

\bm{The problem to be solved is the design of the dynamic controller in the form \eqref{eq:controller_ac}-\eqref{eq:integraldynamics} for the $i$th agent of a bidirectionally interconnected system \eqref{eq:sys} that ensures disturbance string stability and thus \eqref{eq:estimatedss} is satisfied.}

\section{Controller Design} \label{sec:Controller}
The control objective is to drive the states of the system \eqref{eq:sys} to the desired configuration $x_i^\star$, while satisfying the estimate \eqref{eq:estimatedss}, that is disturbance string stability is ensured.

\subsection{Sufficient Conditions for String Stability}
Consider the system \eqref{eq:sys} in closed loop with the controller \eqref{eq:controller_ac} and $k=0$ (no integral action), then a set of sufficient conditions for string stability are \cite{Monteil2019}: 

{\setlength{\abovedisplayskip}{0.5em}
\setlength{\belowdisplayskip}{0.5em}
\setlength{\abovedisplayshortskip}{0.5em}
\setlength{\belowdisplayshortskip}{0.5em}
\begin{description}
	\item[C1] $ h_{i,i-1}(t,x^\star_{i},x^\star_{i-1}) = 0$, $h_{i,i+1}(t,x^\star_{i},x^\star_{i+1}) = 0$, and $h^{0}_{i}(t,x^\star_i,x_0) = 0$;
	\item[C2] for some $c\neq 0$ and $b>0$ \begin{align}
	\begin{split}
	&\mu_2\left(
	\dfrac{\partial f_i(x_i)}{\partial x_i}
	+ \dfrac{\partial u_i}{\partial x_i}
	\right)\leq -c^2, \\
	&\max\left\{\left\lVert
	\dfrac{\partial h_{i,i-1}}{\partial x_{i-1}}
	\right\rVert_2,\left\lVert
	\dfrac{\partial h_{i,i+1}}{\partial x_{i+1}}
	\right\rVert_2\right\}\leq b, \\
	& \text{for all } x_i, x_{i-1}, x_{i+1} \in \mathbb{R}^{2n};
	\end{split}
	\end{align}
	\item[C3] $\varepsilon_i < \frac{c^2}{b}-1$.
\end{description}
}

The following proposition formalise the result in \cite{Monteil2019}.
\begin{prop}[Sufficient Conditions for DSS]
	Assume that the coupling functions $h_{i,j}$ in \eqref{eq:controller_ac} are designed such that conditions \textbf{C1}, \textbf{C2}, and \textbf{C3} are satisfied. Then, the trajectories of the system \eqref{eq:sys} in closed loop with the controller \eqref{eq:controller_ac}, with $k=0$, satisfy
		\begin{equation}\label{eq:prevestimatedss}
			\begin{split}
			\sup_i\left\lvert x_i(t)-x_i^\star(t) \right\rvert_2 &\leq e^{-\bar{c}^2t}\sup_i\left\lvert x_i(0)-x_i^\star(0) \right\rvert_2 \\&+ \frac{1-e^{-\bar{c}^2t}}{\bar{c}^2}\sup_i\left\lVert d_i(t) \right\rVert_{\infty}
			\end{split}
		\end{equation} where $\bar{c}^2 = c^2-b(1+\max_i\varepsilon_i)$, which ensures DSS.
	\label{proposition_monteil}
\end{prop}
\begin{pf}	
	It follows directly from \cite[Theorem 1]{Monteil2019}. ~ \hfill ~ \qed
\end{pf}

The sufficient conditions \textbf{C1}, \textbf{C2}, and \textbf{C3} in Proposition \ref{proposition_monteil} provide a tool to select controllers that ensure string stability of an interconnected system. The proposition also ensures that the states are ultimately bounded for bounded disturbances. However, under the classical scenario of constant disturbances, when a platoon of vehicles hits a sloping road for instance, the states will not converge to the desired values and the state error will not vanish. Moreover, the error will be bounded by the maximum value of the disturbance weighted by an increasing function of time. This behaviour is undesirable and the controller should be able to compensate, at least, for constant disturbances. We propose to design a controller with the integral action capable of rejecting constant disturbances and preserving the DSS property.

\subsection{String Stable Control with Constant Disturbance Rejection}
We augment the system \eqref{eq:sys} with the state $\xi_i \in \mathbb{R}^n$ and consider the controller \eqref{eq:controller_ac}-\eqref{eq:integraldynamics}. Then, the dynamics of the closed loop can be written as follows
\begin{equation}
	\dot{z}_i = \phi_i(z_i) + v_i + \bm{p_i}
	\label{eq:sysclC},
\end{equation}
where $z_i = [x_{i}^T \; \xi^T_i]^T$ is the state vector and $\xi_i = \zeta_i+k^{-1}\bar{w}_i$. Also, we define $v_i = [ 0_{1\times n} \; v_{x,i}^T \; v_{\zeta,i}^T ]^T$, where $0_{a\times b}$ is the $a$-by-$b$ matrix of zeros and $a,b\in\mathbb{N}$, with $v_{x,i} = h_{i,i-1} + \varepsilon_i h_{i,i+1} + h^{0}_{i} + k\xi_i $ and $v_{\zeta,i} =	g_{i,i-1} + \varepsilon_i g_{i,i+1} + g^{0}_{i}$, and the vector \bm{$p_i = [ 0_{1\times n} \; w_i^T \; 0_{1\times n} ]^T$} contains the time varying disturbance.   The system dynamics is defined by $\phi_i\colon \mathbb{R}^{3n} \to \mathbb{R}^{3n}$. Then, the reference for the new state vector becomes $z_i^\star = [{x^\star_i}^T \; 0_{1\times n}]^T$. Note that we dropped the function dependencies to simplify the notation.

We aim at designing $v_i$ such that the closed loop system \eqref{eq:sysclC} is disturbance string stable. Moreover, if conditions \textbf{C1}, \textbf{C2}, and \textbf{C3}  (applied to \eqref{eq:sysclC}) are satisfied, the inclusion of $\xi_i \in \mathbb{R}^n$ and the change of coordinates lead the controller to reject constant disturbances whilst guarantying disturbance string stability. However, directly satisfying those conditions is in general a difficult task.

Similar to \cite{Monteil2019}, we consider a heterogeneous car platoon system where $\phi_i(z_i)=Fz_i$ and we design a controller $v_i$ that meets the control objectives. The system dynamics can be written as
\begin{equation}
	\dot{z}_i = Fz_i + \bm{\frac{1}{m_i}}v_i + \bm{\frac{1}{m_i}p_i}
\label{eq:sysclV},
\end{equation}
with $
	F = [
		0_{3n\times n}~
		[I_{n}~0_{n \times 2n}]^T~
		0_{3n\times n}
		]
$, where $I_{n}$ is the $n$-by-$n$ identity matrix, \bm{and $m_i$ is the mass of vehicle $i$}. It is  convenient to write 
\bm{$v_i = m_i\left[H_{i,i-1} + \varepsilon_iH_{i,i+1} + H^{0}_{i}\right]$}, with and $H_{i,i-1} = [0_{1\times n} \; h_{i,i-1}^T \; g_{i,i-1}^T]^T$, $H_{i,i+1} = [0_{1\times n}\; h_{i,i+1}^T \; g_{i,i+1}^T]^T$, and $ H^{0}_{i} = [0_{1\times n} \; [{h^{0}_{i}} + k\xi_i]^T \; {g^{0}_{i}}^T]^T$.

\bm{Now we modify the sufficient conditions \textbf{C1}, \textbf{C2}, and \textbf{C3} through their application to a linear transformation of the closed loop system \eqref{eq:sysclV}. This facilitates finding the controller gains that satisfy the conditions by posing them as constraints of an optimisation problem.}

\begin{prop}[Transformed Sufficient Conditions]\label{proposition_cond}
	Consider the system \eqref{eq:sysclV} in closed loop with the controller \eqref{eq:controller_ac}-\eqref{eq:integraldynamics}. Let $T_i$ be the transformation matrix, with the coupling constant matrices $\alpha_i \in \mathbb{R}^{n\times n}$ and $\beta_i \in \mathbb{R}^{n\times n}$
	\begin{equation}
		T_i \triangleq \begin{bmatrix}
		I_{n} ~&~ \alpha_i ~&~ 0_{n\times n} \\
		0_{n\times n} ~&~ I_{n} ~&~ \beta_i \\
		0_{n\times n} ~&~ 0_{n\times n} ~&~ I_{n}
		\end{bmatrix}.
		\label{eq:transform}
	\end{equation}
	Assume the following sufficient conditions are satisfied,
	{\setlength{\abovedisplayskip}{0.5em}
\setlength{\belowdisplayskip}{0.5em}
\setlength{\abovedisplayshortskip}{0.5em}
\setlength{\belowdisplayshortskip}{0.5em}
	\begin{description}
		\item[C1*]
		$H_{i,i-1}(t,z^\star_{i-1},z_i^\star) = 0$, $H_{i,i+1}(t,z^\star_{i+1},z_i^\star) = 0$, and $H^{0}_{i}(t,z_i^\star,z_0) = 0$;
		\item[C2*] for some $c\neq 0$ and $b>0$ 
		\begin{align}
		\begin{split}
		&\mu_2\left({J}_{i,i}\right)\leq -c^2, \\
		&\max\left\{\left\lVert {J}_{i,i-1}		\right\rVert_2,\left\lVert {J}_{i,i+1}\right\rVert_2\right\}\leq b, \\
		&\text{for all } z_i,z_{i-1},z_{i+1} \in \mathbb{R}^{3n};
		\end{split}
		\end{align}
		\item[C3*] $\varepsilon_i < \frac{c^2}{b}-1$.
	\end{description}}
	where the Jacobian $J$ of the closed loop system \eqref{eq:sysclV}, is given by the matrices ${J}_{i,i}(\alpha_i,\beta_i,{z}_i,{z}_{i-1},{z}_{i+1},\varepsilon_i)$, ${J}_{i,i-1}(\alpha_i,\beta_i,{z}_i,{z}_{i-1})$ and ${J}_{i,i+1}(\alpha_i,\beta_i,{z}_i,{z}_{i+1})$ below
	\begin{align}
	\begin{split}
	    {J}_{i,i}&= T_iFT_i^{-1} \\
    	&+ 	\begin{bmatrix}
		    	\dfrac{\partial \tilde{v}_i}{\partial z_i}\begin{bmatrix}1 \\ 0 \\ 0\end{bmatrix} & \dfrac{\partial \tilde{v}_i}{\partial z_i} \begin{bmatrix}-\alpha_i \\ 1 \\ 0\end{bmatrix} & \dfrac{\partial \tilde{v}_i}{\partial z_i} \begin{bmatrix}\alpha_i \beta_i \\ -\beta_i \\ 1\end{bmatrix}
	    	\end{bmatrix} \\ 	
		{J}_{i,i\pm 1}&= \begin{bmatrix}
		\dfrac{\partial\tilde{H}}{\partial z}\begin{bmatrix}1 \\ 0 \\ 0\end{bmatrix} & \dfrac{\partial\tilde{H}}{\partial z}\begin{bmatrix}-\alpha_i \\ 1 \\ 0\end{bmatrix} & \dfrac{\partial\tilde{H}}{\partial z}\begin{bmatrix}\alpha_i \beta_i \\ -\beta_i \\ 1\end{bmatrix}
		\end{bmatrix},
	\end{split}
	\label{eq:Jacobian}
	\end{align}
	where $\dfrac{\partial\tilde{H}}{\partial z} \triangleq \dfrac{\partial T_iH_{i\pm 1}}{\partial z_{i\pm 1}}$ and $\tilde{v}_i \triangleq T_iv_i$.
	
	Define $\tilde{z}_i\triangleq T_iz_i$ and $\tilde{p}_i \triangleq T_ip_i$. Then, the following properties hold true.
	
	\emph{(i)} The dynamics of the transformed system are 
	\begin{align}
		\dot{\tilde{z}}_i 
		&=T_iFT_i^{-1}\tilde{z}_i+\tilde{v}_i + \tilde{p}_i
		\label{eq:sysclT}
	\end{align}	
	with the new state vector $\tilde{z}_i$ defined as
	
	\begin{align}
		\tilde{z}_i  &=
		\begin{bmatrix}
		x_{1i} + \alpha_i x_{2i} \\
		x_{2i} + \beta_i \xi_i \\ 
		\xi_i
		\end{bmatrix} = 
		\begin{bmatrix}
		\tilde{z}_i^1 \\
		\tilde{z}_i^2 \\
		\tilde{z}_i^3
		\end{bmatrix}. \label{eq:statesT}
	\end{align}	
	The transformed unperturbed closed loop dynamics are
	\begin{equation}
		\begin{split}
		\dot{\tilde{z}}_i &= T_iFT_i^{-1}\tilde{z}_i +
		T_i[H_{i,i-1}(t,T_i^{-1}\tilde{z}^\star_i,T_{i-1}^{-1}\tilde{z}^\star_{i-1}) \\&+ \varepsilon_iH_{i,i+1}(t,T_i^{-1}\tilde{z}^\star_i,T_{i+1}^{-1}\tilde{z}^\star_{i+1}) + H^{0}_{i}(t,T_i^{-1}\tilde{z}^\star_{i},{z}_0)]
		\end{split}
\raisetag{2.5\normalbaselineskip}
		\label{eq:unperturbedT}
	\end{equation}
	and the transformed desired configuration is $\tilde{z}_i^\star=T_iz_i^\star$.

	\emph{(ii)} The following estimate is satisfied    
	\begin{equation}
		\begin{split}
		\sup_i\left\lvert \tilde{z}_i(t)-	\tilde{z}_i^\star(t) \right\rvert_2 &\leq e^{-\bar{c}^2t}\sup_i\left\lvert \tilde{z}_i(0)-	\tilde{z}_i^\star(0) \right\rvert_2  \\&+ \frac{1-e^{-\bar{c}^2t}}{\bar{c}^2}\sup_i\left\lVert \tilde{m}_i(t) \right\rVert_{\infty}.
		\end{split}
		\label{eq:result_T}
	\end{equation}
\end{prop}
\begin{pf}
First note that \eqref{eq:statesT} follows from simple algebra and \eqref{eq:sysclT} follows by substituting $z_i=T^{-1}\tilde{z}_i$ into \eqref{eq:sysclV}, which proves property (i).

To prove property (ii), we show that the conditions \textbf{C1*}, \textbf{C2*} and \textbf{C3*} are verified if and only if the conditions \textbf{C1}, \textbf{C2} and \textbf{C3} are verified for the dynamics \eqref{eq:unperturbedT}, \bm{which include the integral state $\zeta_i$ and dynamic controller \eqref{eq:controller_ac}-\eqref{eq:integraldynamics}}. Thus, by Proposition~\ref{proposition_monteil}, the closed loop system in coordinates $\tilde{z}_i$ is DSS.
\bm{Notice that \textbf{C1*} results from applying \textbf{C1} to the transformed closed loop dynamics \eqref{eq:unperturbedT}}. Hence, condition \textbf{C1} is satisfied \bm{for the transformed closed loop dynamics} if and only if \textbf{C1*} is satisfied. Now, to compute \textbf{C2}, we differentiate the transformed closed loop dynamics \eqref{eq:sysclT}, and obtain the Jacobian $\tilde{J}\in \mathbb{R}^{3nN\times3nN}$, defined by the matrices $\tilde{J}_{i,i}\in\mathbb{R}^{3n\times3n}$, $\tilde{J}_{i,i-1}\in\mathbb{R}^{3n\times3n}$ and $\tilde{J}_{i,i+1}\in\mathbb{R}^{3n\times3n}$ defined below
\begin{equation}
	\tilde{J}_{i,i} = \frac{\partial (T_iF\tilde{z}_iT_i^{-1})}{\partial \tilde{z}_i} + \frac{\partial \tilde{v}_i}{\partial \tilde{z}_i}, \quad
	\tilde{J}_{i,i\pm 1} = \frac{\partial T_iH_{i\pm1}}{\partial \tilde{z}_{i\pm 1}},
\end{equation}
which can be expressed in terms of the state vector $z_i$ of \eqref{eq:sysclV}, as ${\partial (T_iF\tilde{z}_iT_i^{-1})}/{\partial \tilde{z}_i} = T_iFT_i^{-1}$ and 
{\renewcommand{\arraystretch}{2.0}
\begin{align*}
	 \begin{split}
	\dfrac{\partial \tilde{v}_i}{\partial \tilde{z}_{i}} &= \begin{bmatrix} \dfrac{\partial \tilde{v}_i}{\partial \tilde{z}_i^1} & 
	\dfrac{\partial\tilde{v}_i}{\partial\tilde{z}_i^2} & 
	\dfrac{\partial \tilde{v}_i}{\partial \tilde{z}_i^3}
	\end{bmatrix} = \begin{bmatrix}
	\dfrac{\partial \tilde{v}_i}{\partial z_i} 
	\dfrac{\partial z_i}{\partial \tilde{z}_i^1} & 
	\dfrac{\partial \tilde{v}_i}{\partial z_i} 
	\dfrac{\partial z_i}{\partial \tilde{z}_i^2} & 
	\dfrac{\partial \tilde{v}_i}{\partial z_i} 
	\dfrac{\partial z_i}{\partial \tilde{z}_i^3}
	\end{bmatrix}
	\end{split} 
\end{align*}}
{\renewcommand{\arraystretch}{2.0}
\begin{align*}
	\begin{split}
	\dfrac{\partial T_iH_{i\pm 1}}{\partial \tilde{z}_{i \pm 1}} &= \begin{bmatrix} \dfrac{\partial T_iH_i}{\partial \tilde{z}_{i \pm 1}^1} & 
	\dfrac{\partial T_iH_i}{\partial\tilde{z}_{i \pm 1}^2} & 
	\dfrac{\partial T_iH_i}{\partial \tilde{z}_{i \pm 1}^3}
	\end{bmatrix} \\
	&= \begin{bmatrix}
	\dfrac{\partial T_iH_i}{\partial z_{i \pm 1}} 
	\dfrac{\partial z_{i \pm 1}}{\partial \tilde{z}_{i \pm 1}^1} & 
	\dfrac{\partial T_iH_i}{\partial z_{i \pm 1}} 
	\dfrac{\partial z_{i \pm 1}}{\partial \tilde{z}_{i \pm 1}^2} & 
	\dfrac{\partial T_iH_i}{\partial z_i} 
	\dfrac{\partial z_{i \pm 1}}{\partial \tilde{z}_{i \pm 1}^3}
	\end{bmatrix}.
	\end{split}
\end{align*}
}\clearpage

By solving the partial derivatives above, we obtain the Jacobian matrix $J$ in \eqref{eq:Jacobian}, which is written in terms of the state vector $z_i$, obtaining also condition \textbf{C2*}. This implies that \textbf{C2} is satisfied \bm{for the transformed closed loop dynamics} if and only if \textbf{C2*} is satisfied. Condition \textbf{C3*} follows directly from \textbf{C3}. Finally, the inequality \eqref{eq:result_T} results by application of Proposition \ref{proposition_monteil}. ~ \hfill ~ \qed
\end{pf}
It is important to note that to use the sufficient conditions \textbf{C1*}, \textbf{C2*}, and \textbf{C3*}, it is not necessary to compute the transformed system, but only the Jacobian $J$ and the transformation matrix $T_i$. Also, the change of coordinates and the transformation allows us to find controllers independent of the constant disturbance $\bar{w}_i$. Now we show that if the transformed system \eqref{eq:sysclT} is disturbance string stable, so is the system \eqref{eq:sysclV}.

\begin{prop}[DSS of the Augmented System]
	Consider the system \eqref{eq:sysclV} in closed loop with the controller \eqref{eq:controller_ac}-\eqref{eq:integraldynamics}, and assume that the functions $H_{ij}$ are such that the conditions \textbf{C1*}, \textbf{C2*}, and \textbf{C3*} are satisfied. Then,
	\begin{equation}
	\begin{split}
	\sup_i\left\lvert z(t)-z^\star(t) \right\rvert_2 &\leq Ke^{-\bar{c}^2t}\sup_i\left\lvert z(0)-z^\star(0) \right\rvert_2 \\ &+ K\frac{1-e^{-\bar{c}^2t}}{\bar{c}^2}\sup_i\left\lVert w_i(t) \right\rVert_{\infty}
	\end{split}
	\label{eq:result2}
	\end{equation}
	where $\bar{c}^2 = c^2-b(1+\max_i\varepsilon_i)$ and $K=\frac{\max_i\{\sigma_{\max}(T_i)\}}{\min_i\{\sigma_{\min}(T_i)\}}$.
	\label{proposition_main}	
\end{prop}

\begin{pf}
	Under the assumption, the conditions in Proposition \ref{proposition_cond} are satisfied, and thus \eqref{eq:result_T} holds true. \bm{Then, we define $A_i \triangleq T_i^TT_i$ and $\sigma(T_i) = \sqrt{\lambda(A_i)}$ to obtain the following bound for the quadratic form $z_i^TA_iz_i$}%
	\begin{align}
		\begin{split}
		\underline{\sigma}\sup_i\left\lvert z_i(t)-z_i^\star(t) \right\rvert_2 &\leq \sup_i\left\lvert \tilde{z}_i(t)-	\tilde{z}_i^\star(t) \right\rvert_2 \\
		\sup_i\left\lvert \tilde{z}_i(0)-	\tilde{z}_i^\star(0) \right\rvert_2 &\leq \bar{\sigma} \sup_i\left\lvert z_i(0)-z_i^\star(0) \right\rvert_2 \\
		\sup_i\left\lVert \tilde{p}_i(t) \right\rVert_{\infty} &\leq \bar{\sigma}\left\lVert p_i(t) \right\rVert_{\infty}
		\label{eq:bounds}
		\end{split}
	\end{align}
	where $\bar{\sigma} = \max_i\{\sigma_{\max}(T_i)\}$ and $\underline{\sigma} = \min_i\{\sigma_{\min}(T_i)\}$.
	
	Hence, by using \eqref{eq:bounds} in \eqref{eq:result_T} and noting that $\sup_i\|{p_i(t)}\|_{\infty} = \sup_i\left\lVert w_i(t) \right\rVert_{\infty}$, we obtain
	\begin{equation*}
	\begin{split}
	 \sup_i\left\lvert z_i(t)-z_i^\star(t) \right\rvert_2 &\leq \frac{\bar{\sigma}}{\underline{\sigma}} e^{-\bar{c}^2t}\sup_i\left\lvert z_i(0)-z_i^\star(0) \right\rvert_2 \\&+ \frac{\bar{\sigma}}{\underline{\sigma}}
	 \frac{1-e^{-\bar{c}^2t}}{\bar{c}^2}\left\lVert w_i(t) \right\rVert_{\infty} 
	\end{split}
	\end{equation*}
from where we obtain \eqref{eq:result2} by setting $K=\frac{\bar{\sigma}}{\underline{\sigma}}$. ~ \hfill ~ \qed
\end{pf}

Proposition \ref{proposition_cond} shows DSS of the closed loop dynamics \eqref{eq:sysclV}, which includes the integral action. Now, we will show that DSS holds for the original states of the system \eqref{eq:sys}.

\begin{cor}[DSS of the Original System]
	Consider the system \eqref{eq:sys} in closed loop with the controller \eqref{eq:controller_ac}-\eqref{eq:integraldynamics}. If the functions $H_{ij}$	are such that the conditions \textbf{C1*}, \textbf{C2*}, and \textbf{C3*} are satisfied, then the state errors can be bounded as follows
	\begin{equation}
	\begin{split}
	\sup_i\left\lvert x_i(t)-x_i^\star(t) \right\rvert_2 &\leq  Ke^{-\bar{c}^2t} \sup_i\left\lvert x_i(0)-x_i^\star(0) \right\rvert_2 \\&+ Ke^{-\bar{c}^2t}\sup_i\left\lvert \zeta_i(0)+k^{-1}\bar{w}_i \right\rvert_2  \\&+	 K\frac{1-e^{-\bar{c}^2t}}{\bar{c}^2}\sup_i\left\lVert w_i(t) \right\rVert_{\infty}
	\end{split}
\raisetag{2\normalbaselineskip}
	\label{eq:resultC}
	\end{equation}
	where $\bar{c}^2 = c^2-b(1+\max_i\varepsilon_i)$ and $K=\frac{\max_i\{\sigma_{\max}(T_i)\}}{\min_i\{\sigma_{\min}(T_i)\}}$.
	\label{corollary_org}
\end{cor}

\begin{pf}
	First note that $z_i = [x_i^T \; \xi_i^T]^T$ implies that
	\begin{equation}\label{eq:bound1}
		\sup_i\left\lvert x_i(t)-x_i^\star(t) \right\rvert_2 \leq \sup_i\left\lvert z_i(t)-z_i^\star(t) \right\rvert_2.
	\end{equation}
	As all the assumptions of Proposition \ref{proposition_main} are satisfied, inequality \eqref{eq:result2} holds. Thus, using \eqref{eq:result2} in \eqref{eq:bound1}, we obtain 
	\begin{equation}
		\begin{split}
		\sup_i\left\lvert x_i(t)-x_i^\star(t) \right\rvert_2 &\leq \sup_i\left\lvert z_i(t)-z_i^\star(t) \right\rvert_2 \\
		&\leq Ke^{-\bar{c}^2t}\sup_i\left\lvert z_i(0)-z_i^\star(0) \right\rvert_2 \\ &+ K\frac{1-e^{-\bar{c}^2t}}{\bar{c}^2}\sup_i\left\lVert w_i(t) \right\rVert_{\infty}.
		\end{split}
		\label{eq:bound2}
	\end{equation}
	Also, using the triangle inequality, we can write $\sup_i\left\lvert z_i(0) - z_i^\star(0) \right\rvert_2 \leq \sup_i\left\lvert x_i(0) - x_i^\star(0) \right\rvert_2 + \sup_i\left\lvert \xi_i(0) \right\rvert_2$. Then from \eqref{eq:bound2} and the fact that $\xi_i(0) = \zeta_i(0) + k^{-1}\bar{w}_i$, we obtain \eqref{eq:resultC}, which completes the proof. ~ \hfill ~ \qed
\end{pf} 

Corollary \ref{corollary_org} proves that the state errors of the original system \eqref{eq:sys} are bounded by the initial errors,the initial integral action deviations from their respective equilibria, and by the infinite norm of the time-variant disturbances. Also, the integral action ensures that when the agents are subject to constant disturbances, the states converge to their desired values. The bound \eqref{eq:resultC} is comparable to the bound in Corollary 1 of \cite{Monteil2019}, which is
	\begin{equation} \label{eq:boundcormonteil}
		\begin{split}
		\sup_i&\left\lvert x_i(t)-x_i^\star(t) \right\rvert_2 \leq  K e^{-\bar{c}^2t} \sup_i\left\lvert x_i(0)-x_i^\star(0) \right\rvert_2  \\&+ K\frac{1-e^{-\bar{c}^2t}}{\bar{c}^2}\sup_i\left\lVert \bar{w}_i+w_i(t) \right\rVert_{\infty}.
		\end{split}
        \raisetag{2\normalbaselineskip}
	\end{equation}

\section{Simulations}
\label{sec:Results} 
In this section, we consider $x_i = [q_i \; \dot{q}_i]^T$, where $q_i, \dot{q_i} \in \mathbb{R}$ are the position and speed of vehicle $i$ in a vehicle platoon, whose closed loop dynamics is described by system \eqref{eq:sys}, with $n=1$, $f_{1i}=x_{2i}$ and $f_{2i} = 0$, and controller \eqref{eq:controller_ac}-\eqref{eq:integraldynamics}, with the coupling functions below
\begin{align}
	\begin{split}
	h_{i,i-1} &= h^p_i(q_{i-1}-q_i-\delta_{i,i-1}) + K^v_i(\dot{q}_{i-1}-\dot{q}_i) \\
	h_{i,i+1} &= h^p_i(q_{i+1}-q_i+\delta_{i+1,i}) + K^v_i(\dot{q}_{i+1}-\dot{q}_i) \\
	h^{0}_{i} &= K^{p0}_{i}(q_{0}-q_i-\delta_{i,0}) + K^{v0}_{i}(\dot{q}_0-\dot{q}_i)
	\end{split}
\end{align} 
where $\delta_{i,j}$ is the desired spacing, while $\delta_{i,0} = \sum_{j=1}^{i} \delta_{j,j-1}$ is the distance to the reference $x_0$. The functions that shape the integral action dynamics are
\begin{align}
	\begin{split}
	g_{i,i-1} &= g^p_{i}(q_{i-1}-q_i-\delta_{i,i-1}) + G^v_{i}(\dot{q}_{i-1}-\dot{q}_i)\\
	g_{i,i+1} &= g^p_{i}(q_{i+1}-q_i+\delta_{i+1,i}) + G^v_{i}(\dot{q}_{i+1}-\dot{q}_i)\\
	g^{0}_{i} &= G^{q0}_{i}({q}_0-{q}_i-\delta_{i,0}) + G^{v0}_{i}(\dot{q}_0-\dot{q}_i)
	\end{split}
\end{align}
where we selected $h^p_i(x) = K^p_{1i}\tanh(K^p_{2i}x)$ and $g^p_i(x) = G^p_{1i}\tanh(G^p_{2i}x)$. The controller gains are $K^p_{1i}$, $K^p_{2i}$, $K^v_i$, $K^{p0}_{i}$, $K^{v0}_{i}$, $k$, $G^p_{1i}$, $G^p_{2i}$, $G^v_i$, $G^{p0}_{i}$, and $G^{v0}_{i}$. We compute the partial derivatives in \eqref{eq:Jacobian} and using CVX, a package for specifying and solving convex programs \cite{cvx}, find the controller gains that satisfy the conditions \textbf{C1*}, \textbf{C2*} and \textbf{C3*} so that \eqref{eq:result2} holds. The controller gains are 
are $\alpha_i = 0.3$,
       $\beta_i = -0.4$,
       $\varepsilon_i=1$,
       $  K^p_{1i} = K^p_{2i} = 0.1188$,
       $  K^v_i = 0.0121$,
       $ K^{p0}_i = 0.6$,
       $ K^{v0}_i = 0.6$,
       $    k = 0.2508$,
       $  G^p_i = 0.01$,
       $  G^v_i = 0.01$,
       $ G^{p0}_i = 0.2881$, and
       $ G^{v0}_i = 0.3420$.

\bm{We consider a string of $N=5$ vehicles} and compare the controllers $C_1$ and $C_2$ obtained using Corollary 1 in \cite{Monteil2019} and Proposition \ref{proposition_main} respectively. \bm{The initial conditions are $x_i(0)=[q_0(0)-\delta_{i,0} + r, \dot{q}_0(0)+ r]^{T}$, the inter-vehicle spacing is $\delta_{i,i-1}=\delta_{i+1,i}=10\text{ m}$ and the reference speed $\dot{q}_0=20\text{ m/s}$. We set the time-variant disturbance as $w_i(t)=r \sin(t) \exp(-0.1t)$ and the constant disturbance is $\bar{w}_i = (1 + r)\text{ m/s}^2$, where $r$ is uniformly randomly generated in the interval $[-1,1]$ and updated after each use. We considered the case where the controller uses the nominal vehicle mass $\hat{m}_i = 1000 \text{ kg}$, while the mass in the model is $m_i = \hat{m}_i + 200 r$.}

\begin{figure}
	\centering
	\psfragfig*{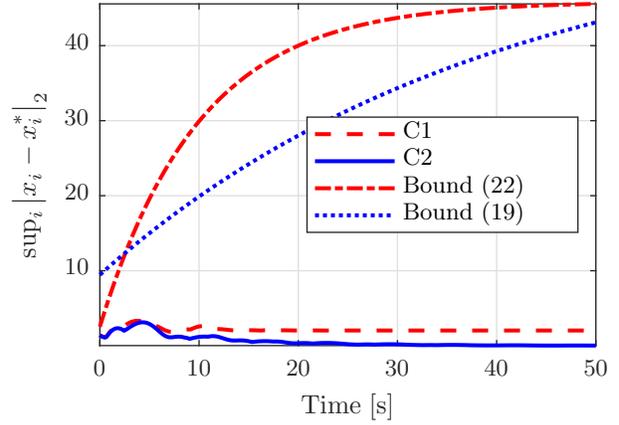}
	\vspace*{-1em}
	\caption{\bm{State errors norm and bounds for $C_1$ and $C_2$.}}
	\label{fig:normsTime}
\end{figure}

\begin{figure}
	\centering
	\psfragfig{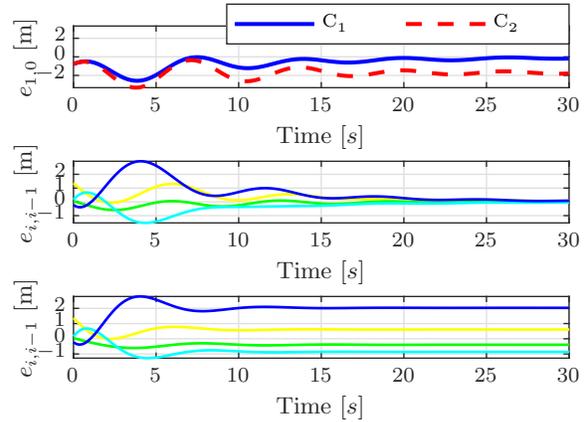}
	\vspace*{-1em}
	\caption{\bm{Inter-vehicle distance error $e_{1,0}$ for $C_1$ and $C_2$ (top), and $e_{i,i-1}$ for $C_2$ (middle) and $C_1$ (bottom).}}
	\label{fig:delta}
\end{figure}

\begin{figure}
	\centering
	\psfragfig{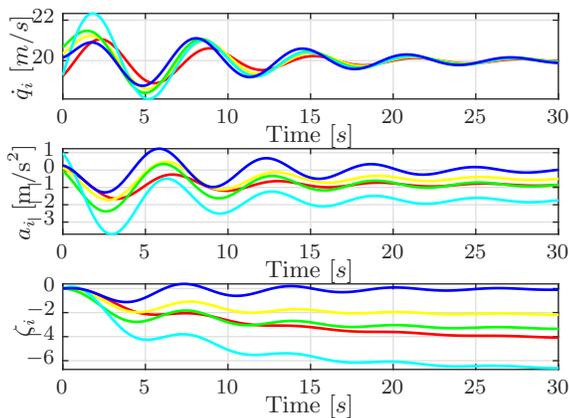}
	\vspace*{-1em}
	\caption{\bm{Velocity (top), acceleration (middle) and integral state (bottom) time histories.}}
	\label{fig:vel_zeta}
\end{figure}

\bm{Figure \ref{fig:normsTime} compares the state error norm and the bounds \eqref{eq:boundcormonteil} and \eqref{eq:resultC}. As expected, since the time-varying disturbances vanish, the state error norm converges to a constant when using the controller $C_1$. However, the state error norm is suppressed when using the controller $C_2$, which shows better performance and compensates for the constant disturbances. We notice that uncertainty in the masses is also compensated by the controller.} Figure \ref{fig:delta} shows the deviation of the inter-vehicle distances, that is $e_{i,i-1}=q_{i-1}-q_{i}-\delta_{i,i-1}$, from the desired value. We note that the controller $C_1$ cannot compensate for disturbances and the inter-vehicle distances do not converge to the desired values, but controller $C_2$ does whilst ensuring a reasonable inter-vehicle distances during the transient. \bm{Figure \ref{fig:vel_zeta} shows that the vehicles achieve the reference velocity. Also, notice that the integral action states converge to a value proportional to the constant disturbance, and the acceleration signals $a_i = u_i/m_i$ are smooth and within reasonable values.}

\section{Conclusion} \label{sec:Conclusion}
In this paper, we presented sufficient conditions that guarantee disturbance string stable for a linear, asymmetric, bidirectional, interconnected system under nonlinear control with integral action. Under these conditions, the controller will ensure the state errors are bounded by functions of the initial conditions and the time-variant (zero mean) disturbance, whilst rejecting constant (non-zero mean) disturbance due to the addition of integral action.
Future work will focus on developing sufficient conditions for controllers that use local information without global knowledge of reference signals.

\begin{ack}
	G.F.S., A.M., and J.F. acknowledge continued support from the Queensland University of Technology (QUT) through the Centre for Robotics. The authors thank Dr. J. Monteil for fruitful discussions and for sharing the controller and optimisation code in \cite{Monteil2019}.
\end{ack}

\bibliographystyle{plain}
\bibliography{references_Aut2020}

\end{document}